# Deep Transform: Time-Domain Audio Error Correction via Probabilistic Re-Synthesis

Andrew J.R. Simpson [#1]

[#] Centre for Vision, Speech and Signal Processing, University of Surrey
Surrey, UK
[1] Andrew.Simpson@surrey.ac.uk

*Abstract*—In the process of recording, storage and transmission of time-domain audio signals, errors may be introduced that are difficult to correct in an unsupervised way. Here, we train a convolutional deep neural network to re-synthesize input time-domain speech signals at its output layer. We then use this abstract transformation, which we call a *deep transform* (DT), to perform probabilistic re-synthesis on further speech (of the same speaker) which has been degraded. Using the convolutive DT, we demonstrate the recovery of speech audio that has been subject to extreme degradation. This approach may be useful for correction of errors in communications devices.

## I. INTRODUCTION

Recording, storage and transmisison of audio signals are fundamental to modern communication, media and culture in general. In the process of recording, storage and transmisison there are many possible errors that may be introduced. For example, distortion may be introduced during recording, digitization or processing, and losses may be introduced during magnetic media storage, network transmission, etc. Therefore, some unsupervised means to correct errors that would exploit prior knowledge about the scope of what is represented by the audio data would be useful.

Deep neural networks [1], [2], [3], [4], [5] (DNN) learn abstract feature representations from data [6], [7]. A DNN that is trained to synthesize its inputs at its output layer is known as an autoencoder [2]. We may think of the autoencoder as an abstract transformation device – a *deep transform* (DT) – that embodies abstract knowledge of the features of the data [7]. The DT may therefore be used to transform new data through the learned abstract feature space, constituting an abstract filter [7].

In a previous paper [7], it was shown that an autoencoder DNN could be used for probabilistic re-synthesis of degraded images. In that study, image classification performance was used as metric to demonstrate error correction. Here, we extend the same approach to the correction of errors in time-domain audio signals and we use an objective source separation metric [8] to quantify recovery from error.

We trained a convolutive DNN autoencoder on frames of speech audio data that were free of errors and then used the resulting model as a transformation device (i.e., a convolutive DT). We then used this convolutive DT to probabilistically re-synthesize new audio data that had been degraded by having a random proportion replaced with random data. We demonstrate the recovery of time domain audio data from serious errors using the convolutive DT to perform probabilistic re-synthesis.

## II. METHOD

We consider the problem of speech audio that has been arbitrarily degraded by non-additive (nonlinear) errors [7]. We chose a long segment of continuous speech from a single (male) speaker. The speaker read a passage from a book and was recorded in pseudo anechoic conditions. The time domain (monaural) speech signal was divided into two portions. The first two minutes of speech were used as training data, and the next ten seconds were used as test data. The speech audio was decimated to a sample rate of 4 kHz and divided up into overlapping frames of length 1000 samples. The overlap interval was 10 samples for the training data and 1 sample for the test data. This gave approximately 50,000 frames of training data and approximately 40,000 frames of test data.

To obtain a DT, we trained an autoencoder DNN on the training frames of audio. We then used the DT to re-synthesize each degraded (test) frame a number of times, each time with further random errors added, and averaged the result. Finally, to capture the degree of recovery from error, we computed the signal-to-distortion ratio (SDR) for the recovered audio signal using the BSS-EVAL toolkit [8].

The input layer to our network was a vector of length 1000. The speech audio signal was normalized to unit scale and mean of 0.5. This allowed us to use a signmoidal output layer that is mapped to the range [0,1]. The network used the biased sigmoid activation function [6] throughout and bias was set to zero for the output layer. The network was of size 1000x2500x1000 units and was trained to replicate its own inputs at its output layer. The autoencoder was trained on the ~50,000 training examples using stochastic gradient descent (SGD). The model was trained for 600 iterations, with a relatively slow learning rate. Each iteration of SGD consisted of a complete sweep of the entire training data set. The model was trained without dropout.

*Degradation.* Prior to division into overlapping windows for feedforward processing using the autoencoder DT, the test audio was subjected to nonlinear degradation. A certain proportion of the ~40,000 samples was randomly chosen and replaced with a random value selected from a distribution of equivalent mean and standard deviation to the audio. The test audio was degraded at levels between 0 and 100% (of the samples) at intervals of 5. Fig. 1a provides a waveform representing the test audio degraded to a level of 10 % random replacement. Note that this is not the same as additive noise.

*Convolutive Probabilistic Re-Synthesis via DT.* The degraded test audio was then divided into overlapping frames. Each frame (length 1000 samples) of the degraded test audio was transformed using the autoencoder a number ($N$) of times. Prior to each of the $N$ separate transforms, 50% of the samples of the degraded frame (chosen at random) were replaced with random values. I.e., each time the respective degraded frame was further degraded, but these secondary degradations were different for each transform. The activation of the output layer of the autoencoder was taken as re-synthesized audio. The resulting distribution, of $N$ re-synthesized instances of each degraded audio frame, was then averaged to provide an error-corrected frame. In order to account for neurons in the output layer that were invariantly active, all error-corrected frames (i.e., of the entire test set) were then averaged and the result subtracted from each frame. The frames were then superposed in a sliding-window fashion, the results averaged and the DC offset removed (zero mean was restored).

The degraded test audio and the error-corrected test audio were then compared with the original test audio by computing the SDR using the BSS-EVAL toolkit [8] for the whole 10-second audio signal. The SDR gives an overall, objective measure of distortion (error) with respect to the original signal.

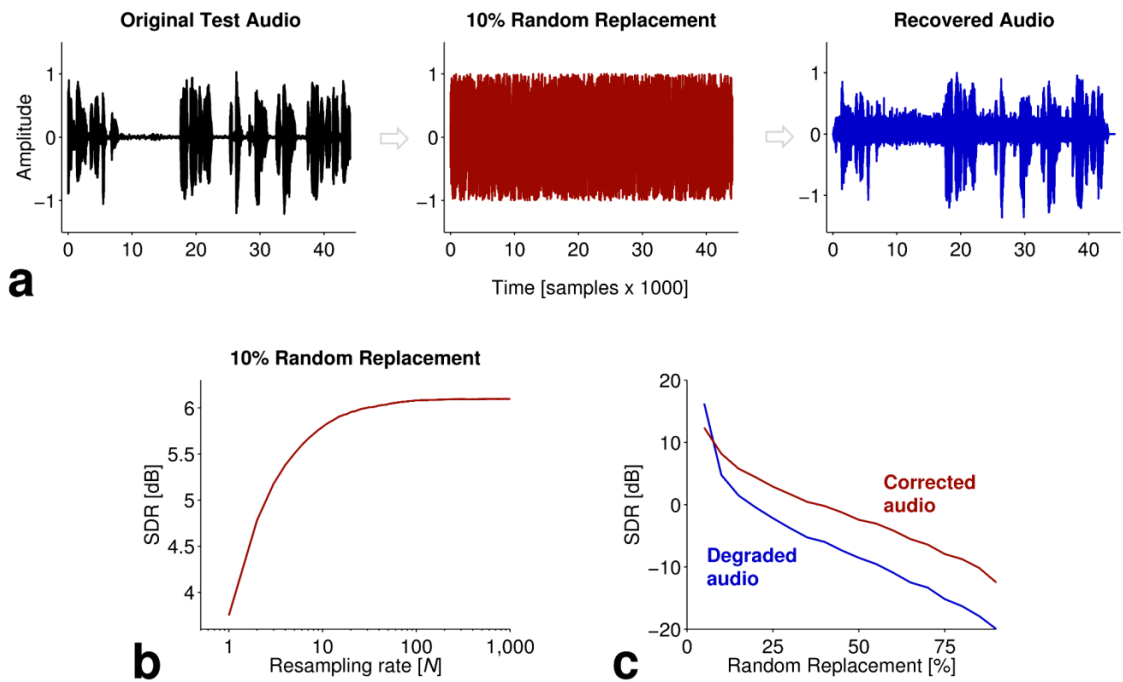

**Fig. 1. Probabilistic audio re-synthesis via convolutive deep transform. a** shows waveforms of the original test audio (left), the degraded audio (middle; 10% degradation) and the recovered audio corrected by DT probabilistic re-synthesis (right; resampling rate of $N$ = 1000). **b** Signal-to-distortion ratio (SDR) as a function of re-sampling ($N$) rate for test audio degraded at 10% random degraded (random replacement) test audio. **c** SDR as a function of degradatation for both the degraded audio (blue) and the corrected audio (red).

III. RESULTS

Fig. 1a plots waveforms for the test audio before (left) and after degradation (middle) and recovery (right). The middle plot shows the effect of 10% noise replacement. The right most plot demonstrates recovery via DT probabilistic re-synthesis ($N$ = 1000). The recovered audio signal appears true to the overall features of the original, including the essential envelope and asymmetric features.

Fig. 1b plots SDR as a function of resampling rate ($N$) for the test audio degraded to a level of 10%. This plot demonstrates the cumulative effect of re-sampling on probabilistic re-synthesis. The function tends towards the asymptote around 6 dB. For reference, the SDR of this specific instance of the degraded audio (10% random replacement), prior to any recovery, is 1.9 dB. Thus, the overall improvement in SDR is around 4 dB. The fact that the SDR function starts around 3.8 dB (at $N$ = 1) is the result of the overlapping windows of the convolutive process. In other words, even for a resampling rate of $N = 1$, because each overlapping windowed frame is re-sampled independently, in the act of superposition of the overlapping windows there is an inherent (and additional) resampling rate that corresponds to the window size and overlap interval.

Fig. 1c plots SDR as a function of degradation from 5% to 95% degradation (random replacement). The blue line plots the SDR of the degraded audio and the red line plots the SDR for the signal recovered via DT probabilistic re-synthesis with a resampling rate of $N = 100$. The degree of recovery is proportional to the degree of degradation ($r = 0.98$, $P < 0.01$, *Pearson Product-Moment*) and peaks at around 9 dB improvement at a degradation rate of 95%. In summary, we have demonstrated recovery of audio from errors via convolutive DT probabilistic re-synthesis.

## IV. DISCUSSION AND CONCLUSION

In this paper, we have demonstated a convolutive DNN autoencoder employed as an abstract transformation device – a *convolutive deep transform* (CDT). We have shown that this CDT can be used for probabilistic re-synthesis of degraded time-domain audio and that this procedure allows speech audio to be recovered from extreme error. This point was illustrated by improved SDR for the degraded speech audio recovered via CDT probabilistic re-synthesis. Random replacement with noise represents a relatively extreme form of error and recovery of the original signal by other means (such as manual audio editing tools) would be arduous or even impossible.

This unsupervised approach to audio error correction might be useful for mobile telecommunications, either generalized to speech or tailored to the speech of specific (i.e., known) speakers. The approach may also be useful for forensic analysis or recovery of audio. For example, given a suitable 'clean' training example of speech from a given speaker, it would (in principle) be possible to recover degraded speech from a different recording of the same speaker. More generally, as in the previous examples [7], the outcome of the DT probabilistic re-synthesis demonstrated here resembles the perceptual error correction facilities of the brain. Hence, our findings may provide insight into how the brain might exploit noisy representations for the purposes of error correction.

ACKNOWLEDGMENT

AJRS did this work on the weekends and was supported by his wife and children.